\documentclass[aps,pra,twocolumn,groupeaddress,showkeys]{revtex4-1}
\usepackage[utf8]{inputenc} 
\usepackage{amsmath}
\usepackage{braket}
\usepackage{graphicx}
\usepackage{pgfplots}
\usepackage{csquotes}
\usepackage{hhline}
\usepackage{amssymb}

\usepackage{dcolumn}
\usepackage{tabularx}
\usepackage[colorlinks=true,linkcolor=red,citecolor=magenta,urlcolor=blue]{hyperref}
\usepackage{longtable}
\usepackage{braket}
\usepackage{float}
\newcolumntype{C}{>{\centering\arraybackslash}X}

\begin{document}


\title{Solving Linear Systems of Equations by Using the Concept of Grover's Search Algorithm: An IBM Quantum Experience}

\author{Rituparna Maji}
\email{rituparnamaji98@gmail.com}
\affiliation{Department of Physics, Central University of Karnataka, Karnataka 585367, India}%

\author{Bikash K. Behera}
\email{bkb18rs025@iiserkol.ac.in}
\author{Prasanta K. Panigrahi}
\email{pprasanta@iiserkol.ac.in}
\affiliation{Department of Physical Sciences, Indian Institute of Science Education and Research Kolkata, Mohanpur 741246, West Bengal, India}

\begin{abstract}
Quantum algorithm, as compared to classical algorithm, plays a notable role in solving linear systems of equations with an exponential speedup. Here, we demonstrate a method for solving a particular system of equations by using the concept of well-known Grover's quantum search algorithm. The algorithm finds the solution by rotating the initial state vector in the Hilbert space to get the target solution state. It mainly involves finding particular matrices that solve the set of equations and constructing corresponding quantum circuits using the basic quantum gates. We explicitly illustrate the whole process by taking 48 different set of equations and solving them by using the concept of Grover's algorithm. We propose new quantum circuits for each set of equations and design those on the IBM quantum simulator. We run the quantum circuit for one set of equations and obtain the desired results, and hence verify the working of the algorithm.
\end{abstract}

\maketitle

\section{Introduction}
It is well known that a quantum computer has superior power than a classical one to perform various computational tasks \cite{qge_NCBook} due to the harnessing of quantum mechanics. Grover's search algorithm \cite{qge_GrovSTOC96}, Shor's algorithm \cite{qge_ShorSIAM99} and Deutsch-Jozsa algorithm \cite{qge_DJPRSA92} have significant contributions in demonstrating the advantage of quantumness over their classical counterparts. Quantum computers are used to simulate quantum mechanical problems \cite{qge_FIJTP82,qge_ADSLS97,qge_ZWL11,qge_ADLH05,qge_LBPNC10} which show exponential speedups. 

Linear equations can be used to describe a number of phenomena that occur in nature. Hence, solving linear systems of equations becomes an integral part to understand the way the mysteries of nature unfold. Linear equations are used vastly in all the fields of science and engineering. Calculating the solutions for N linear equations in N unknowns requires a time scale of $O(N)$ through a classical computer. However, it is evident that a quantum computer consumes a logarithmic time scale order to approximately compute the value of a function which can serve the same purpose. Thus it is useful to use the quantum algorithm to achieve exponential speedups for large value of N. Harrow \emph{et al.} have come up with a way to solve a set of linear equations \cite{qge_HarrowPRL2009}. Here, we propose a way to solve a particular set of linear equations by the usage of Grover's search algorithm. 

The basic idea for solving the problem is described as follows. We are given a N $\times$N matrix (A), and a unit vector ($\vec{y}$) such that the following relation holds, $A\vec{x} = \vec{y}$. Our main goal is to find the solution vector ($\vec{x}$) which satisfies the above relation. Initially, we make initial assumptions that, the unit vectors ($\vec{x}, \vec{y}$) are represented as quantum states of a system of appropriate number of qubits. The solution ($\vec{x}$) can be easily obtained if we can find another unitary matrix U such that UA is an identity matrix and hence, $\vec{x}=U\vec{y}$. That means, we can prepare $\vec{y}$ using appropriate number of qubits and elementary gates, then we can design the UA operation and applying on $\vec{y}$, we can get $\vec{x}$. This is the way, here we solve the particular set of equations, where the matrix A describes how the linear combination of $\vec{x}$ should be there to get $\vec{y}$. It is to be noted that, in our proposed algorithm, we only consider a particular set of equations where the elements of matrix A contain either $1/2$ or $-1/2$. Having the above elements we can have 48 matrices corresponding to 48 set of equations. We propose new quantum circuits for solving each of the set of equations, design and simulate 8 quantum circuits, and verify the results using the IBM quantum experience platform.  

The quantum processors developed by IBM Quantum Experience have been used for performing various quantum tasks  (see \cite{qge_RundlePRA2017,qge_GrimaldiSD2001,qge_GangopadhyayQIP2018,qge_MajumderarXiv2017,qge_LiQMQM2017,qge_SisodiaQIP2017,qge_VishnuQIP2018,qge_GedikPRA2017,qge_DasharXiv2017,qge_RoyarXiv2017,qge_SatyajitQIP2018,qge_SolanoQMQM2017,qge_WoottonQST2017,qge_BertaNJP2016,qge_DeffnerHel2017}). Foundational work in quantum information theory such as Leggett-Garg \cite{qge_HuffmanPRA2017}, Mermin inequality \cite{qge_AlsinaPRA2016,qge_GarciaJAMP2018}, nohiding theorem \cite{qge_KalraQIP2019} and Hardy's paradox \cite{qge_DasarXiv2017} have been tested. Useful applications e.g., quantum cheque \cite{qge_MoulickQIP2016,qge_BeheraQIP2017}, topological quantum walks \cite{qge_BaluQST2018}, quantum permutation algorithm \cite{qge_YalcinkayaPRA2017}, quantum error correction \cite{qge_WoottonPRA2018,qge_GhoshQIP2018}, estimation of molecular ground state energy \cite{qge_KandalaNAT2017}, quantum artificial life \cite{qge_RodriguezSciRep2018}, quantum machine learning \cite{qge_SchuldEPL2017,qge_DuttaarXiv2018}, quantum repeater \cite{qge_BeheraQIP2019}, quantum tunneling \cite{qge_HegadearXiv2017,qge_MalikRG2019}, quantum robots \cite{qge_MahantiQIP2019} have been demonstrated using the quantum processors. 

The rest of the paper is organized as follows. Sec. \ref{qge_Sec2} discusses Grover's search algorithm and its geometrical visualization. Sec. \ref{qge_Sec3} describes our proposed procedure to solve the given linear system of equations. Finally, we conclude in Sec. \ref{qge_Sec4} by discussing future directions of our work.

\section{Grover's search algorithm \label{qge_Sec2}}
In 1997, Grover proposed a quantum search algorithm for searching in an unstructured database \cite{qge_GroverPRL1997}, where it takes only $\sqrt{N}$ steps for finding an object as compared to the classical algorithm that takes N/2 steps. It is the fastest way to transform a vector from its initial state to a final desired state. The algorithm finds the solution by rotating the initial state vector in the Hilbert space in iterative sequences until it gets the target solution state. This is achieved by a repeated application of an operator known as Grover operator. Let us first visualize this algorithm geometrically, as this concept will be later used to propose our algorithm. 

\begin{figure}[H]
\centering
\includegraphics[scale=0.4]{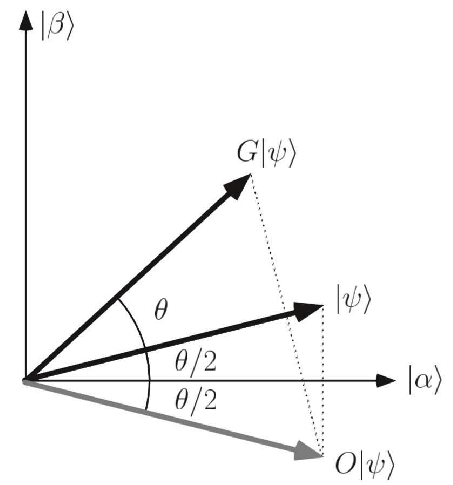}
\caption{Geometric representation of Grover's search algorithm.}
\label{qge_Fig1}
\end{figure}

In Fig. \ref{qge_Fig1}, $\vert\psi\rangle$ is the initial state vector, which is represented as an equal superposition of all the basis vectors which includes $\vert\beta\rangle$, i.e., the solution vector and $\vert\alpha\rangle$, the superposition of all the basis states excluding the solution vector $\vert\beta\rangle$. $\vert\alpha\rangle$ and $\vert\beta\rangle$ form a complete set of basis vectors and hence are orthogonal to each other.

Here, $\theta/2$ is the angle between the initial state $\vert\psi\rangle$ and the state ($|\alpha\rangle$). Now, we apply an oracle operation $U_{f}$ which operating on $\vert\psi\rangle$ rotates the solution state $\vert\beta\rangle$ by an angle $\theta/2$ and acts as an identity to the other states $\vert\alpha\rangle$. Now the another matrix $U_{g}$ rotates the reflected matrix ${\vert\psi\rangle}$ by an angle $\theta$. The product of $U_{f}$ and $U_{g}$ is called the Grover operation ($U=U_{f}$.$U_{g})$). The U matrix rotates the initial state $\vert\psi\rangle$ in this way in iterative sequence until it reaches the target solution state $\vert\beta\rangle$. If the iteration is done k times on $\vert\psi\rangle$, the initial state will be rotated by an angle (2k+1)$\theta$. In order to obtain a state close to the solution state we have to perform Grover's operation k times such that $\sin{(2k+1)\theta} \approx 1$.

To represent a vector in $2^{n}$ dimensional Hilbert space, n qubits are required, where $\vert\psi\rangle$ will be an equal superposition of $2^n$ computational basis states. The solution state, $|\beta\rangle$ is usually an equal superposition of M computational basis states where $0<M<N$. After certain approximations, it is evident that the number of times the Grover iteration has to be repeated in order to get arbitrarily close to $\vert\beta\rangle$ is $O(\sqrt{N/M})$.

\section{The oracle \label{qge_Sec3}}

\begin{figure*}
    \centering
	\includegraphics[width=\textwidth]{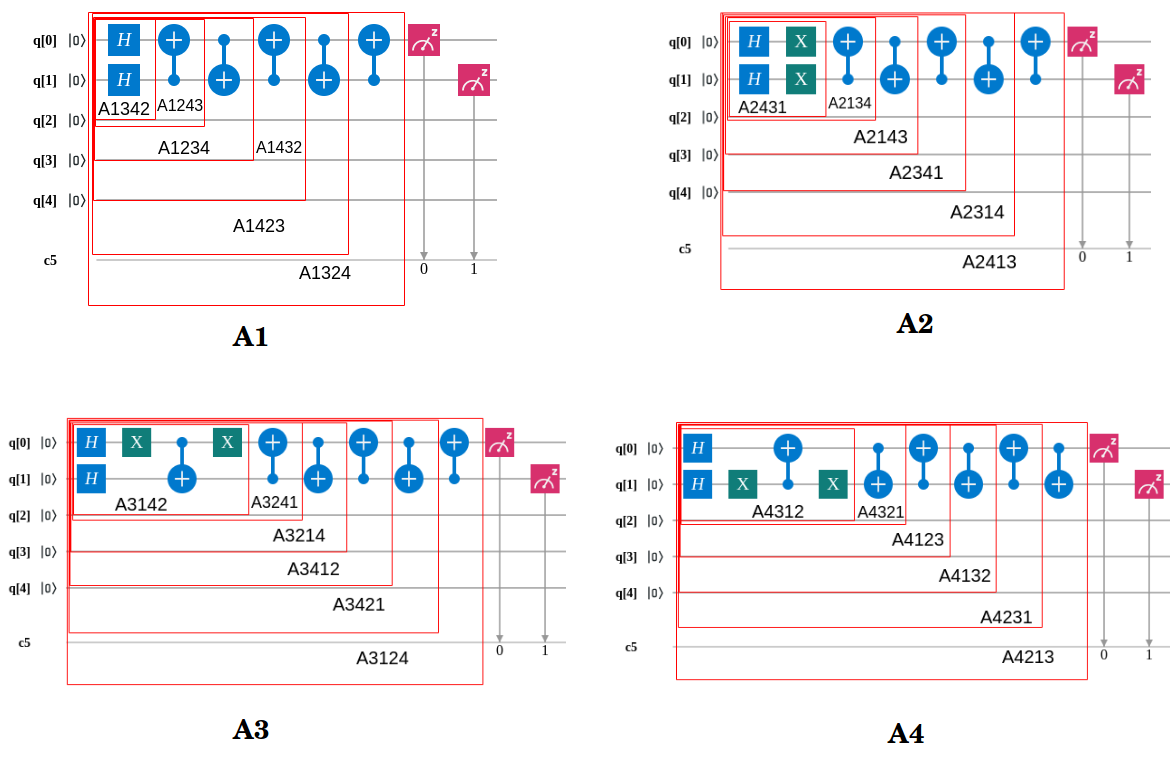}
    \caption{The 4 circuits A1, A2, A3, A4 each containing 6 different quantum circuits separated by red boxes. Each quantum circuit when applied on initial state $\vec{y}$ gives the solution state $\vec{x}$ for the corresponding set of equations labelled by their corresponding matrix names.}
    \label{qge_Fig2}
\end{figure*}

\begin{figure*}
    \centering
	\includegraphics[width=\textwidth]{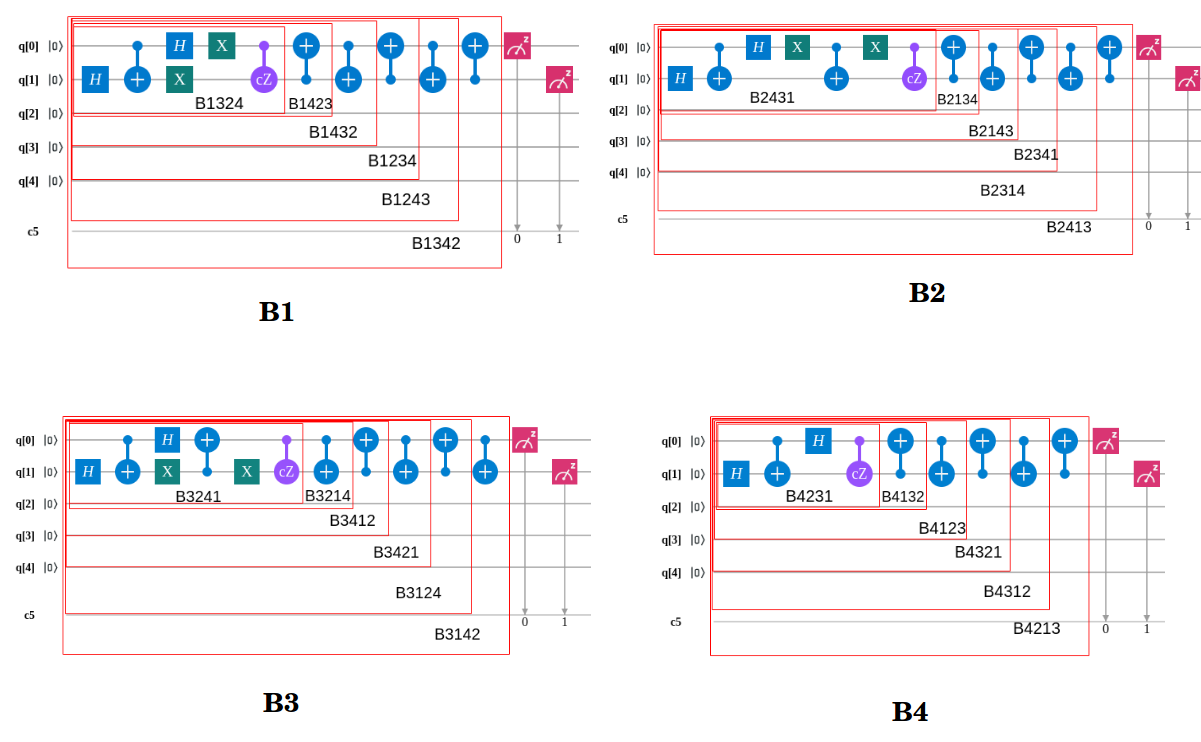}
    \caption{The 4 circuits B1, B2, B3, and B4 each containing 6 different quantum circuits separated by red boxes. Each quantum circuit when applied on the initial state $\vec{y}$ gives the solution state $\vec{x}$ for the corresponding set of equations labelled by their corresponding matrix names.}
    \label{qge_Fig3}
\end{figure*}

As mentioned in the introduction section, we need to find a unitary operator U such that UA is an identity operation and consequently, $\vec{x}=U\vec{y}$. We construct the unitary operator in such a way that when it applies on A matrix, converts the $i^{th}$ column in the A matrix operation, into $i^{th}$ column of the Identity matrix, where one element is being 1 and others are 0, eventually converting a column of all superposition states into a particular target state. Let's consider the A matrix, which follows the below conditions. 

\begin{equation}\sum_{i=1}^{N}(A_{ij})^2= 1\label{qge_Eq1}\end{equation}
where, $A_{ij}$ denotes an element of $i^{th}$ row and $j^{th}$ column of the matrix A. It is to be noted that, $\vert y\rangle$ must also be normalized. i.e.,
\begin{equation}\sum_{i=1}^{N}y_i^2= 1\label{qge_Eq2}\end{equation}
Now, we can represent each column in matrix A as $\ket{\psi_i}$. Hence,
\begin{equation}
A = [\ket{\psi_1} \ket{\psi_2} \ket{\psi_3} ... \ket{\psi_N}]
\label{qge_Eq3}
\end{equation}

After the application of $U$,
\begin{equation}
UA = [U\ket{\psi_1} U\ket{\psi_2} U\ket{\psi_3} ... Un\ket{\psi_N}]=I_{N \times N}
\label{qge_Eq4}
\end{equation}

For all $\vert\psi_i\rangle$, we can find a unitary operation which performs the above operation. It is to be noted that this method of solving linear system of equations using the concept of Grover's algorithm works especially for matrices which follow the above conditions. Here, in our algorithm we find solutions to those matrices whose column matrices have elements $1/2$ or $-1/2$. The same concept can be used to solve other set of equations too.

\subsection{Example}

Let us consider the following set of equations which we will solve to find the solution; 
\begin{eqnarray}
x_1+x_2+x_3+x_4=2\nonumber\\
x_1-x_2-x_3+x_4=0\nonumber\\
x_1-x_2+x_3-x_4=0\nonumber\\
x_1+x_2-x_3-x_4=0
\label{qge_Eq5}
\end{eqnarray}

The above set of equations can be written in the matrix form as follows;
\begin{equation}
A_{1234}\vec{x}=\vec{y}
\label{qge_Eq6}
\end{equation}
where,
\begin{eqnarray}
A_{1234} &=& \frac{1}{2}\left[\begin{array}{cccc}
1&1&1&1\\
1&-1&-1&1\\
1&-1&1&-1\\
1&1&-1&-1\\
\end{array} \right] \nonumber\\
\vec{x}&=&[x_1\hspace{.2cm}
  x_2 \hspace{.2cm}
 x_3 \hspace{.2cm}
 x_4]^t \nonumber \\ \vec{y}&=&[1\hspace{.2cm}
 0\hspace{.2cm}
 0\hspace{.2cm}
 0]^t
\label{qge_Eq7}
\end{eqnarray}

Here, we can find the following unitary operator, 
\begin{eqnarray}
U_{1234} &=& \frac{1}{2}\left[\begin{array}{cccc}
1&1&1&1\\
1&-1&-1&1\\
1&-1&1&-1\\
1&1&-1&-1\\
\end{array} \right]
\label{qge_Eq8}
\end{eqnarray}
which if applied on A, gives identity matrix ($U_{1234}A_{1234}=I$). Thus multiplying U on both the sides of Eq. \eqref{qge_Eq6}, we have,
\begin{equation}
\vec{x}=U_{1234}\vec{y}    
\label{qge_Eq9}
\end{equation}

\begin{figure*}
    \centering
	\includegraphics[width=\textwidth]{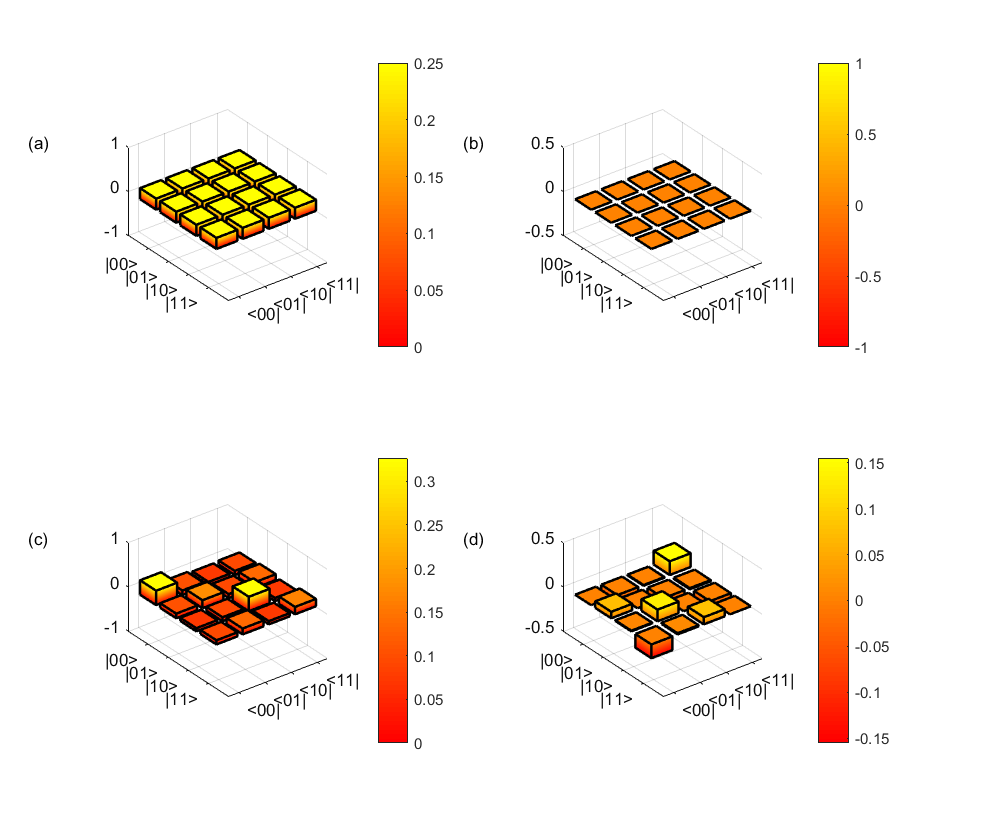}
    \caption{\textbf{Tomography plots for $A_{1234}$ case.} a) and b) represent the reconstructed theoretical density matrices for the solution state. c) and d) represent the reconstructed experimental density matrix for the solution state. The fidelity is found to be 0.9878.}
    \label{qge_Fig4}
\end{figure*}

It is to be noted that, U in this case is coincidentally found to be exactly equal to A, however it might have different form according to A. Now, we explicate the above process in terms of quantum circuits (Fig. \ref{qge_Fig2}, A1). As from Eq. \eqref{qge_Eq9}, it can be seen that applying $U_{1234}$ on $\vec{y}$, we can get our solution vector $\vec{x}$, hence we first prepare the initial state $\vec{y}=[1\hspace{.2cm}
0\hspace{.2cm}
0\hspace{.2cm}
0]^t$ taking two qubits in $|00\rangle$ state. Then we decompose the above unitary operator into two Hadamard gates and two CNOT gates as seen in Fig. \ref{qge_Fig2}, A1 (see the circuit labelled by $A_{1234}$). Now, we apply the unitary operator $U_{1234}$ on the initial state $\vec{y}$, and from the probabilities, after taking the square root of those we can have, the solution vector $\vec{x}$. It can be noted that as we are only dealing with real numbers, hence it will not be a problem here considering the square root of probability as the probability amplitude. The quantum circuit is run on the 5-qubit quantum chip and the state tomography is plotted. Fig. \ref{qge_Fig4} illustrates the tomography plots for the solution vector corresponding to the $A_{1234}$ matrix. The solution state is prepared with 0.9878 fidelity. 

Similarly, for all 48 set of equations, the quantum circuits are designed for constructing their corresponding unitary operators. The 48 set of equations are divided into 8 categories where they share common gates required to build the unitary operators. Hence, they are characterized into 8 sets, and each set contains 6 quantum circuits corresponding to each set of equations. The Figs. \ref{qge_Fig2} and \ref{qge_Fig3} illustrate all the quantum circuits and their corresponding matrices for each set of equations. As any A matrix can have 6 variants by fixing the first column and changing the rest columns, hence 8 matrices have their each 6 variants which are shown in the above figures.  

$A_{1234} = \left[\ket{\psi_1} \ket{\psi_2} \ket{\psi_3} \ket{\psi_4}]\right]$. By doing permutation between the columns of the matrix $A_{1234}$, we can get $4!=24$ different matrices which can be further discussed in 4 subset $(A_{1}, A_{2}, A_{3}, A_{4})$ each containing 6 matrices keeping its first column fixed and permuting between the other 3 columns like given below:\\ $A_{1}:\{A_{1234}$, $A_{1243}$, $A_{1324}$, $A_{1342}$, $A_{1423}$, $A_{1432}\}$, $A_{2}$:$\{A_{2134}$, $A_{2143}$, $A_{2314}$, $A_{2341}$, $A_{2413}$, $A_{2431}\}$, $A_{3}:\{A_{3124}$, $A_{3142}$, $A_{3214}$, $A_{3241}$, $A_{3412}$, $A_{3421}\}$. $A_{4}:\{A_{4123}$, $A_{4132}$, $A_{4213}$, $A_{4231}$, $A_{4312}$, $A_{4321}\}$. Similarly the B matrices are divided into B1, B2, B3 and B4 classes. 

For the above 48 matrices representing 48 different set of equations, we have designed 48 different quantum circuits and discussed them in 8 sets, each containing 6 different quantum circuits. Set $A_{1}$ contains 6 different circuits separated by the red boxes and labeled as their corresponding matrices. Though measurement gate is shown only at the last, but readers are advised to understand that there are measurement gates after each red boxes. To find the solution for the matrix $A_{1234}$, we need to use the circuit $A_{1234}$ followed by the measurement gates. Similarly, it works for all 48 set of quantum circuits representing the corresponding 48 set of equations. 

We simulate the above 8 cases (A1 to A4 and B1 to B4) for 8 different set of equations by designing the quantum circuits on the IBM quantum simulator and obtain the required probabilities. The simulated results are shown in Table \ref{qge_Tab1}. 

\begin{table}[h]
	\centering
	\caption{The Experimental Result}
	\begin{tabular}{|c|c|c|c|c|c}
		\hline
		\textbf{circuit name} & \multicolumn{4}{c|}{Probability amplitude for the state} \\ 
		\hline\textbf{    }&\textbf{0000} & \textbf{0001} & \textbf{0010} & \textbf{0011}\\
		\hline
		$A_{1324}$  &21.875\% & 24.805\%&27.051\%&26.27\%\\
		\hline
				$A_{2413}$ &24.023\% & 25.781\%&23.926\%&26.27\%\\
		\hline
				$A_{3124}$ &24.414\% & 24.609\%&25.684\%&25.293\%\\
		\hline
			$A_{4213}$ &24.316\% & 24.121\%&26.563\%&25\%\\
		\hline
			$B_{1342}$ &24.707\% & 24.832\%&24.219\%&23.242\%\\
		\hline
				$B_{2413}$ &24.902\% & 26.66\%&23.047\%&25.391\%\\
		\hline
			$B_{3142}$ &24.805\% & 25.977\%&24.707\%&24.512\%\\
		\hline
				$B_{4213
			}$  &26.758\% & 25.098\%&25.195\%&22.949\%\\
				\hline

	\end{tabular}
	\label{qge_Tab1}
\end{table}

In the present work, we have used the concept of Grover's algorithm to design the unitary operator for a particular set of equations, where we have taken only equal superposition states in the matrix elements. If we take any set of any arbitrary linear superposition states in the matrix elements, then the same concept can be used, i.e., we need to find an unitary operator which can take from an initial state to a target state for all the coloumns of the matrices corresponding to that particular set of equations. 

\section{Conclusion \label{qge_Sec4}}
To conclude, we have illustrated an approach for solving a particular system of linear equations by using the concept of Grover's search algorithm. We have explicitly taken an example and designed the equivalent quantum circuit for the set of equation. We run the quantum circuit on the 5-qubit real quantum chip, ``ibmqx4" and obtain the solution state with 0.9878 fidelity. We have verified the experimental results for other set of equations by simulating the quantum circuits on the quantum simulator, and hence our approach is found to be successful. In future, this work can be extended for solving more general linear system of equations. 

\section*{Acknowledgements} B.K.B. is financially supported by IISER-K Institute fellowship. R.M. acknowledges all her teachers for sharing their knowledge and also for their continuous support and guidance and her family for their love and moral support. R.M. also acknowledges the hospitality provided by IISER Kolkata during the project work. We are extremely grateful to IBM team and IBM Quantum Experience project.

\end{document}